\documentclass[amsmath,amssymb,amsfont,aps,prl,superscriptaddress,twocolumn, notitlepage, nobibnotes,nofootinbib,longbib,showkeys, 10pt]{revtex4-2}

\usepackage{bm}
\usepackage{xcolor}
\usepackage{physics}
\usepackage{soul}
\usepackage{graphicx}
\usepackage[colorlinks,citecolor=blue,linkcolor=blue,urlcolor=blue,breaklinks=true]{hyperref}

\graphicspath{{./figures/}}

\definecolor{THc}{rgb}{0.7,0.3,0.2}

\definecolor{wyc}{rgb}{0.6,0.0,0.6}

\begin{document}

\title{A scalable advantage in multi-photon quantum machine learning}

\author{Yong Wang}
\thanks{These authors contributed equally to this work. {Current affiliation: University of Oxford, Department of Statistics, United Kingdom}}
\email{yong.wang@stats.ox.ac.uk}
\affiliation{University of Vienna, Faculty of Physics, Vienna Center for Quantum
Science and Technology (VCQ), Boltzmanngasse 5, Vienna A-1090, Austria}

\author{Zhenghao Yin}
\thanks{These authors contributed equally to this work.}
\email{zhenghao.yin@univie.ac.at}
\affiliation{University of Vienna, Faculty of Physics, Vienna Center for Quantum
Science and Technology (VCQ), Boltzmanngasse 5, Vienna A-1090, Austria}
\affiliation{University of Vienna, Faculty of Physics, Vienna Doctoral School of Physics (VDSP), Boltzmanngasse 5, Vienna A-1090, Austria}

\author{Tobias Haug}
\thanks{These authors contributed equally to this work.}
\email{tobias.haug@u.nus.edu}
\affiliation{Quantum Research Center, Technology Innovation Institute, Abu Dhabi, UAE}

\author{Ciro Pentangelo}
\affiliation{Ephos Srl, Viale Decumano 34, 20157 Milano, Italy}

\author{Simone Piacentini}
\thanks{Current affiliation: Quandela SAS, 10 Boulevard Thomas Gobert, 91120 Palaiseau, France}
\affiliation{Istituto di Fotonica e Nanotecnologie, Consiglio Nazionale delle Ricerche (IFN-CNR), piazza L. Da Vinci 32, 20133 Milano, Italy}

\author{Andrea Crespi}
\affiliation{Dipartimento di Fisica, Politecnico di Milano, piazza L. Da Vinci 32, 20133 Milano, Italy}
\affiliation{Istituto di Fotonica e Nanotecnologie, Consiglio Nazionale delle Ricerche (IFN-CNR), piazza L. Da Vinci 32, 20133 Milano, Italy}

\author{Francesco Ceccarelli}
\affiliation{Istituto di Fotonica e Nanotecnologie, Consiglio Nazionale delle Ricerche (IFN-CNR), piazza L. Da Vinci 32, 20133 Milano, Italy}
\affiliation{Ephos Srl, Viale Decumano 34, 20157 Milano, Italy}

\author{Roberto Osellame}
\affiliation{Istituto di Fotonica e Nanotecnologie, Consiglio Nazionale delle Ricerche (IFN-CNR), piazza L. Da Vinci 32, 20133 Milano, Italy}
\affiliation{Ephos Srl, Viale Decumano 34, 20157 Milano, Italy}

\author{Philip Walther}
\email{philip.walther@univie.ac.at}
\affiliation{University of Vienna, Faculty of Physics, Vienna Center for Quantum Science and Technology (VCQ), Boltzmanngasse 5, Vienna A-1090, Austria}
\affiliation{
Christian Doppler Laboratory for Photonic Quantum Computer, Boltzmanngasse 5, Vienna A-1090, Austria}


\begin{abstract}
Photons are promising candidates for quantum information technology due to their high robustness and long coherence time at room temperature. 
Inspired by the prosperous development of photonic computing techniques, recent research has turned attention to performing quantum machine learning on photonic platforms.
Although photons possess a high-dimensional quantum feature space suitable for computation, a general understanding of how to harness it for learning tasks remains blank.
Here, we establish both theoretically and experimentally a scalable advantage in quantum machine learning with multi-photon states.
Firstly, we prove that the learning capacity of linear optical circuits scales polynomially with the photon number, enabling generalization from smaller training datasets and yielding lower test loss values.
Moreover, we experimentally corroborate these findings through unitary learning and metric learning tasks, by performing online training on a fully programmable photonic integrated platform. 
Our work highlights the potential of photonic quantum machine learning and paves the way for achieving quantum enhancement in practical machine learning applications.
\end{abstract}


\maketitle

\section{Introduction}
\label{sec:intro}

\begin{figure*}[!th]
    \centering
    \includegraphics[width=1\linewidth]{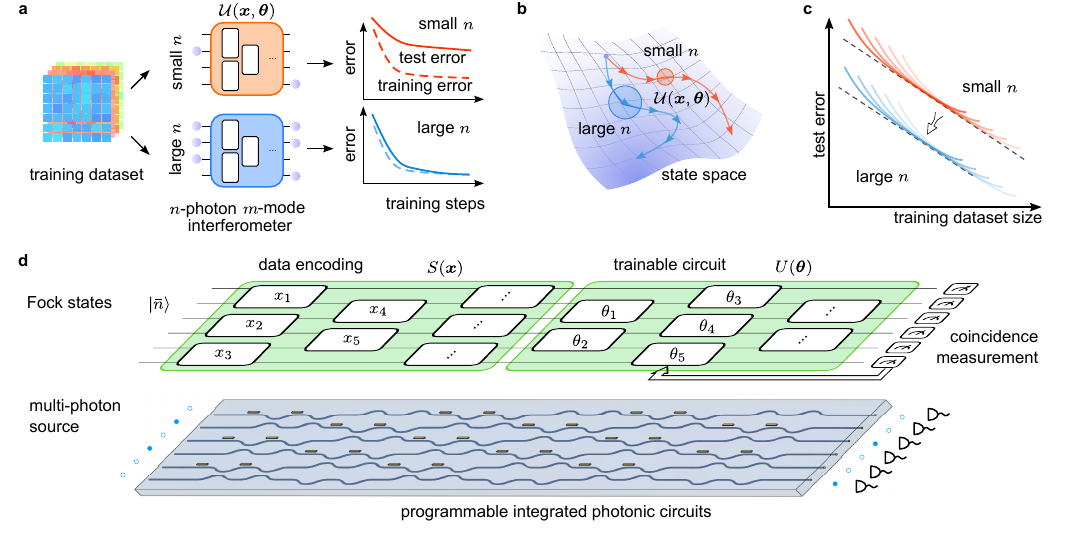}
    \caption{
    \textbf{Architecture of multi-photon quantum machine learning.}
    \textbf{a}. 
    Our model consists of multi-photon states with $n$ photons (shown as dots) propagating through a parameterized linear optical circuit $\mathcal{U}(\vb*{x, \theta})$ with $m$ modes, which encodes feature vectors $\vb*{x}$ and trainable parameters $\vb*{\theta}$. 
    Photons are measured at the output modes, where the number of possible outcomes increases with $n$ and leads to different learning results. 
    \textbf{b}. Illustration of the accessible quantum state space during training with respect to $\vb*{\theta}$. 
    The size of the learnable state space increases with $n$, enhancing the expressivity and trainability of the QML model. 
    \textbf{c}. The scaling trend of a photonic QML model describes the relationship between test error and the training dataset size.
    A higher photon number $n$ corresponds to a larger learning capacity of the model, leading to lower test error and less training data required for learning.
    \textbf{d}. Our multi-photon QML setup using a programmable integrated photonic circuit. 
    Identical photons are generated by multi-photon sources and injected into a photonic chip.
    The chip consists of tunable optical interferometers which imprint either feature vector data $\vb*{x}$ or trainable parameters $\vb*{\theta}$ as phase shifts onto the photon state.
    Each block in the circuit represents a $2\times2$ Mach-Zehnder interferometer containing two tunable phase shifters.
    The final states are measured via coincidence counting of photons in each mode, and used to update the parameters during the learning process.
    } 
    \label{fig:1}
\end{figure*}

The interplay of quantum computing and machine learning, known as quantum machine learning (QML), has emerged as an area of significant research interest in recent years~\cite{biamonte2017quantum, bhartiNoisyIntermediatescaleQuantum2022, preskill2018QuantumComputing}. 
With the anticipation of introducing substantial speedup or enhancement to machine learning problems and data science, certain QML protocols have demonstrated the potential to surpass their classical counterparts, namely reinforcement learning, quantum state learning, and quantum kernel methods~\cite{saggio2021ExperimentalQuantum, huangProvablyEfficientMachine2022, liu2021RigorousRobust}.
Moreover, the noise resilience and ability of learning tasks make QML a suitable candidate for principle validation of quantum computing techniques on near-term noisy quantum hardware~\cite{cerezo2022challenges, preskill2018QuantumComputing}.


As a fundamental quantum particle, photons are characterized by their robustness against decoherence, scalability potential, and room-temperature operability. 
Benefiting from these ideal properties, photonic platforms have been extensively explored and utilized in quantum information processing, including quantum simulation~\cite{ma2011QuantumSimulation, Walther2005} and quantum computing~\cite{Knill2001, OBrien2009, carolan2015universal, maring2024versatile, psiquantumteam2025ManufacturablePlatform}. 
These successes highlight the experimental feasibility of performing QML with photons, with proof-of-pricinple demonstrations including reinforcement learning~\cite{saggio2021ExperimentalQuantum}, quantum-enhanced kernel methods~\cite{yin2025experimental}, variational quantum algorithms~\cite{zhang2022resource, Carolan2020, hoch2025quantum}, and photonic nonlinearity~\cite{Spagnolo}.

Moreover, photons are also theoretically anticipated as a powerful framework for computing tasks.
The classical simulation complexity and potential computational advantage of photonic systems usually depend on the number of photons propagating within the quantum systems.
As the Hilbert space of multi-photon states expands exponentially with the number of photons and modes~\cite{Steinbrecher2019, aaronson2011computational}, specialized tasks such as boson sampling are known to become classically hard for sufficiently high photon numbers~\cite{aaronson2011computational,zhong2020quantum,aaronson2016complexity}.
However, whether scaling up photonic systems can introduce advantages for practical machine learning problems or QML applications has not been explored from both theoretical and experimental perspectives.

In this study, we address this gap by demonstrating a scalable advantage in quantum machine learning with multi-photon states.
We show that the learning capacity of general parametrized linear optical circuits increases polynomially with the photon number, corresponding to an enlarged set of trainable directions in the quantum state space.
We theoretically and experimentally verify that this scalability directly translates into improved learning performance in terms of training dataset size and test error (Fig.~\ref{fig:1}a–c).
Specifically, (1) in unitary learning, multi-photon states reduce the required training data from a linear to a constant scale for generalization; (2) in metric learning, higher test accuracy can be achieved with multi-photon states under the same variational ansatz and training conditions.
The learning processes of these two tasks are realized by online training on a programmable integrated photonic circuit platform~\cite{pentangelo_high-fidelity_2024, crespi2013IntegratedMultimode} based on femtosecond laser waveguide writing~\cite{corrielli2021FemtosecondLaser}, further underscoring the experimental viability and promise of multi-photon QML (Fig.~\ref{fig:1}d).

\section{Multi-photon Enhanced Quantum Machine Learning}
\label{sec:landscape}
\begin{figure*}[!th]
    \centering
    \includegraphics[width=1\linewidth]{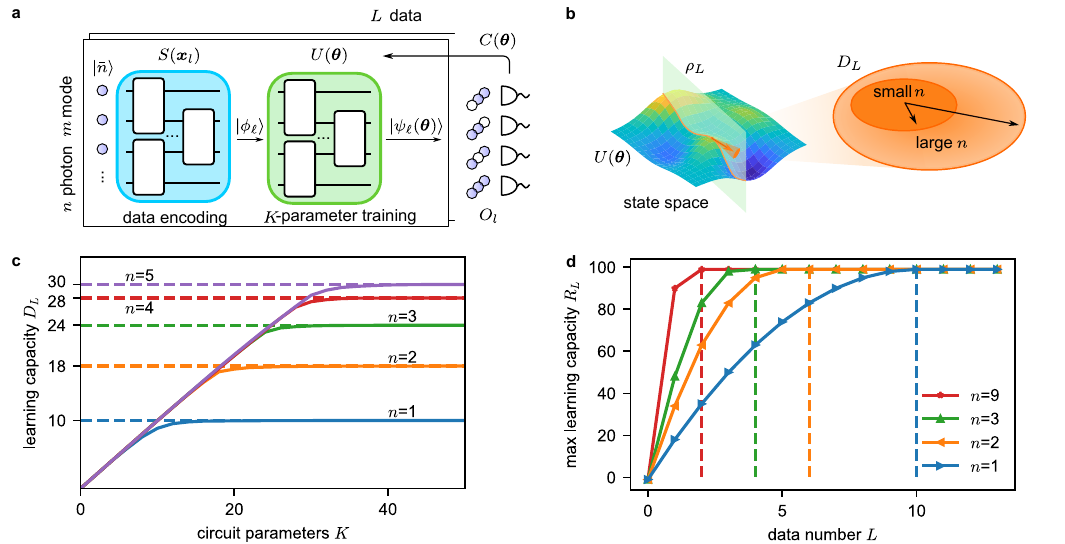}
    \caption{
        \textbf{Learning capacity of linear optical circuits.} 
        \textbf{a}. 
        A photonic QML model is trained on $L$ training data, consisting of $n$-photon states $\{\ket{\phi_\ell}\equiv \ket{\phi(\vb*{x}_\ell)}\}_{\ell=1}^L$ generated by a data-encoding unitary $S(\vb*{x}_\ell)$ carrying feature vector $\vb*{x}_\ell$. 
        A parameterized unitary $U(\vb*{\theta})$ with $K$ trainable circuit parameters $\vb*{\theta}$ is optimized with respect to the loss function $C(\vb*{\theta})$.
        \textbf{b}. 
        Sketch of the state space that can be accessed by $U(\vb*{\theta})$ constrained to $L$ training data. The size of this space is characterized by the learning capacity $D_L(n)$ (the rank of DQFIM $\mathcal{Q}_{ij}$), which increases with $n$ and correlates with learning performance.
        \textbf{c}. Learning capacity $D_L$ as a function of the number of trainable parameters $K$ for different photon numbers $n$ with $m=6$ and $L=1$.
        $D_L$ increases with $K$, reaching the theoretical maximum value predicted by Eq.~\ref{eq:capacity} as horizontal dashed lines.
        \textbf{d}. Maximal learning capacity $R_L(n)$ as a function of the training dataset size $L$ for various $n$ with $m=10$.
        $R_L(n)$ increases with $L$ and saturates at the critical dataset size $L_\text{c}$, indicated by vertical dashed lines. 
    }
    \label{fig:2}
\end{figure*}

Our setup for supervised learning in multi-photon QML is shown in Fig.~\ref{fig:2}a.
We have a set of $L$ different training data $T_L=\{\ket{\phi_\ell},y_\ell\}_{\ell=1}^L$, with each pure training state given by
\begin{equation}
    \ket{\phi_\ell}=S(\vb*{x}_\ell)\ket{\bar{n}}\,,
\end{equation}
where $S(\vb*{x}_\ell)$ is an $m$-mode unitary that encodes the feature vector $\vb*{x}_\ell$, $y_\ell$ is the corresponding label, and $\ket{\bar{n}}$ is an $n$-photon input state~\cite{yin2025experimental}.
Here, we restrict ourselves to Fock states with at most one photon per mode, i.e.,
\begin{equation}
\ket{\bar{n}}=\ket{n_1,n_2,\cdots,n_m},\,n_i\in\{0,1\}\,,
\end{equation}
where $n_i$ denotes the number of photons in the $i$th mode and $n=\sum_{i}n_i$. 
The training data is then processed by a quantum neural network
\begin{equation}
    \ket{\psi_\ell(\vb*{\theta})}=U(\vb*{\theta})\ket{\phi_\ell}\,,
\end{equation}
which is given by a linear optical circuit $U(\vb*{\theta})$ with $K$ trainable circuit parameters $\vb*{\theta}=\{\theta_1, \cdots, \theta_K \}$. 
The quantum neural network is optimized using the training data $T_L$ with respect to a loss function $C(\vb*{\theta})$, and then applied to test data that was not seen during training. 

The power of machine learning models to learn and generalize can be understood from their complexity~\cite{du2021LearnabilityQuantum,abbas2021power,banchi2021generalization,caro2022GeneralizationQuantum,haug2023generalization,gil2024understanding}.
Here, we use a quantum geometric measure, known as the data quantum Fisher information matrix (DQFIM)~\cite{haug2021capacity,haug2023generalization}, which is based on the Fubini-Study metric~\cite{liu2020quantum}. 
Intuitively, the DQFIM quantifies the size of the accessible state space depending on the chosen dataset. 
This identifies the overparameterization transition where training finds good local minima~\cite{larocca2023TheoryOverparametrization}, as well as when sufficient data is available for generalization~\cite{haug2023generalization}.
While previous works were concentrated on qubit-based quantum neural networks, we now apply the DQFIM to multi-photon linear optics. 
It is given by the $K\times K$ matrix~\cite{haug2021capacity, haug2023generalization}
\begin{align}\label{eq:qfim}
    \nonumber
    \mathcal{Q}_{ij} (\vb*{\theta}) &=  4 \Re [  \tr( \partial_i U(\vb*{\theta}) \rho_L \partial_j U^\dagger(\vb*{\theta}) ) \\ 
        & -\tr( \partial_i U(\vb*{\theta}) \rho_L U^\dagger(\vb*{\theta}) ) \tr( U(\vb*{\theta}) \rho_L \partial_j U^\dagger(\vb*{\theta}) ) ]\,.
\end{align}
Here, $\partial_i = \partial / \partial \theta_i$ 
and the mixture of training data $\rho_L = L^{-1} \sum_\ell \rho_\ell$.
The rank of the DQFIM $D_L=\mathrm{rank}(\mathcal{Q}_{ij})$ characterizes the \textit{learning capacity}, i.e., the number of independent directions in the parameter space $\vb*{\theta}$ that can affect the model~\cite{larocca2023TheoryOverparametrization,haug2021capacity,haug2023generalization}. 
As illustrated in Fig.~\ref{fig:2}c, $D_L(K)$ increases with the number of circuit parameters $K$, until saturating at a critical number $K\geq K_\text{c}$. 
At this point, the model becomes overparameterized and achieves the \textit{maximal learning capacity} for a given number of training data $L$
\begin{equation}
    R_L(n)=\max_K \text{rank}(\mathcal{Q}_{ij})\,.
\end{equation}
Such overparameterized models have a good optimization landscape with local minima close to the optimal solution and thus can be easily trained~\cite{larocca2023TheoryOverparametrization}.
Note that $R_L$ is also the upper bound on the rank of the classical Fisher information metric, which characterizes the complexity of neural networks~\cite{abbas2021power}. 

For linear optical circuits, we can compute $R_L(n,m)$ analytically
\begin{align}\label{eq:capacity}
    \nonumber
    R_L(n,m) &\leq 2mnL-n^2L^2-1-(n-1)\delta_{L,1}&(nL\leq m)\,,\\
    R_L(n,m) &= m^2-1 -(m-1)\delta_{L,1}&(nL> m)\,.
\end{align}
where $\delta_{L,1}=1$ for $L=1$ and zero otherwise.
A complete proof is provided in Supplementary Note 2.2. 
We highlight that $R_L(n)$ scales polynomially with photon number $n$.
In particular, for $L=\text{const}$, we have $R_L(n=1)\propto m$, while $R_L(n\propto m)\propto m^2$.
Moreover, $R_L$ is related to the minimal number of training states needed to generalize~\cite{haug2023generalization}, as $R_L$ grows with $L$ and reaches its maximal value at a critical data number $L_\text{c}$. 
For $L\geq L_\text{c}$, the dataset is overcomplete, i.e., one has sufficient data to fully characterize the model. 
In this regime, assuming a noise-free dataset, the model can generalize with low test error~\cite{haug2023generalization}. 
We find that $L_\text{c}$ depends on photon number $n$ as
\begin{equation}\label{eq:lc}
   L_\text{c}(n,m)=\lceil(m-1)/n\rceil+1\,.
\end{equation}

We calculate $R_L$ as a function of $L$ for different values of $n$ in Fig.~\ref{fig:2}d.
A higher photon number yields larger learning capacity $R_L$, leading to a significantly lower number of training states $L_\text{c}$.
In fact, using $n=1$ single-photon states (which also applies to coherent states produced by lasers, see Supplementary Note 2.2 for details), one requires $L_\text{c}=m$ training data to generalize, while $L_\text{c}=2$ for $n=m-1$. 
Therefore, multi-photon states allow for improved learning performance and generalization capability, such as reducing the amount of training data or achieving higher test accuracy.

\section{Learning unitaries with multi-photon states}
\label{sec:unitary}

\begin{figure}
    \centering
    \includegraphics[width=1\linewidth]{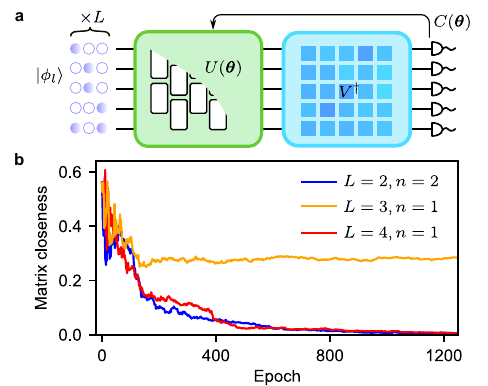}
    \caption{
    \textbf{Experimental unitary learning with multi-photon states.}
    \textbf{a}. We aim to learn an unknown unitary $V$ by training a parameterized unitary $U(\vb*{\theta})$ such that $U(\vb*{\theta})=V$. 
    \textbf{b}. Experimental results of learning a $5\times5$ unitary $V$. 
    We plot the matrix closeness $C_\text{M}(U(\vb*{\theta}),V)$ up to local phases and permutations.
    Perfect generalization corresponds to $C_\text{M}=0$.
    For $n=2$ photons, we achieve nearly optimal closeness $C_\text{M}\approx0$ using only $L=2$ training data (blue curve). 
    In contrast, $n=1$ photon requires at least $L\geq4$ training data (red curve),
    and fails with $C_\text{M}\approx 0.3$ for $L=3$ (orange curve). 
    With one additional training state, we can learn the full unitary $V$ including local phases and mode permutations, as experimentally verified in Supplementary Note 4. 
    }
    \label{fig:3}
\end{figure}

Now, we study the task of variationally learning a black-box unitary~\cite{khatri2019quantum}.
Conventional quantum tomography of quantum process unitaries usually involves complete bases of both input and output states~\cite{obrien2004QuantumProcess,mohseni2008QuantumprocessTomographya}.
Meanwhile, with limited input states, one could use post-selection to learn the unitary adaptively, as demonstrated by variational unsampling and adaptive boson sampling~\cite{Carolan2020, hoch2025quantum}.
Here we show that by using multi-photon states, it is possible to learn the full unitary using only a constant number of training states, without the need of complete bases and post-selection.

We assume that we have oracle access to an unknown $m \times m $ unitary $V$, without any prior knowledge of its internal structure.
The goal is to learn a parameterized unitary $U(\theta)$ such that it represents $V$. 
To do so, we are given $L$ arbitrary $n$-photon training data $\ket{\phi_\ell}$, along with their corresponding labels $V\ket{\phi_\ell}$, forming the training dataset $T_L=\{\ket{\phi_\ell}, V\ket{\phi_\ell}\}_{\ell=1}^L$. 
The parameterized ansatz $U(\vb*{\theta})$ is applied to each training state, project onto the label state $V\ket{\phi_\ell}$, and minimize the loss function $C_\text{train}( \vb*{\theta} ) = 1 - L^{-1}\sum_{\ell=1}^L \abs{ \bra{\phi_\ell} V^\dagger U(\vb*{\theta}) \ket{\phi_\ell} }^2$.
To simplify the presentation, we focus on learning $V$ up to local phase shifts and mode permutations; the full protocol that includes learning these components is described in Supplementary Note 3.

We report the experiment of learning a $5\times5$ unitary with different photon numbers $n$ and training dataset sizes $L$ in Fig.~\ref{fig:3}b. 
Our setup adapts a programmable integrated photonic circuit and an optical switch to implement the learning task for single-photon or two-photon input states.
Pseudo photon-number resolution is used to detect bunching events in the two-photon case.
The parameterized unitary $U(\vb*{\theta})$ is trained from the same random initialization for all datasets by optimizing the parameters of the photonic chip using simultaneous perturbation stochastic approximation (SPSA)~\cite{spall1992multivariate}.
To determine whether the correct unitary has been learned, we compute the matrix closeness $C_\text{M}(U(\vb*{\theta}),V)$, which quantifies the difference between $U(\vb*{\theta})$ and $V$. 
In particular, $C_\text{M}$ equals zero when the two unitaries match up to local phases and mode permutations (see Methods).

We consider three different training datasets: 
one set with size $L=2$ of two-photon states ($n=2$), and another two larger sets with $L=3$ or $L=4$ of single-photon states ($n=1$).
We find that to minimize matrix closeness and thereby achieve generalization, four single-photon training states ($L=4$) are required, while three ($L=3$) are insufficient.
In contrast, two-photon states generalize successfully with only $L=2$, requiring half as much training data as the single-photon case.
This behaviour can be related to the learning capacity $R_L(n)$, which saturates at a critical number of training data that scales approximately as $L_\text{c}\propto 1/n$.
Indeed, the experimentally observed values of $L$ required for generalization have matched our theoretical prediction exactly, confirming that more photons reduce the data needed for generalization (Supplementary Note 3).
We also experimentally demonstrate that the full unitary, including local phases and mode permutations, can be learned with one additional training state (Supplementary Note 4).

\begin{figure*}[!th]
    \centering
    \includegraphics[width=1\linewidth]{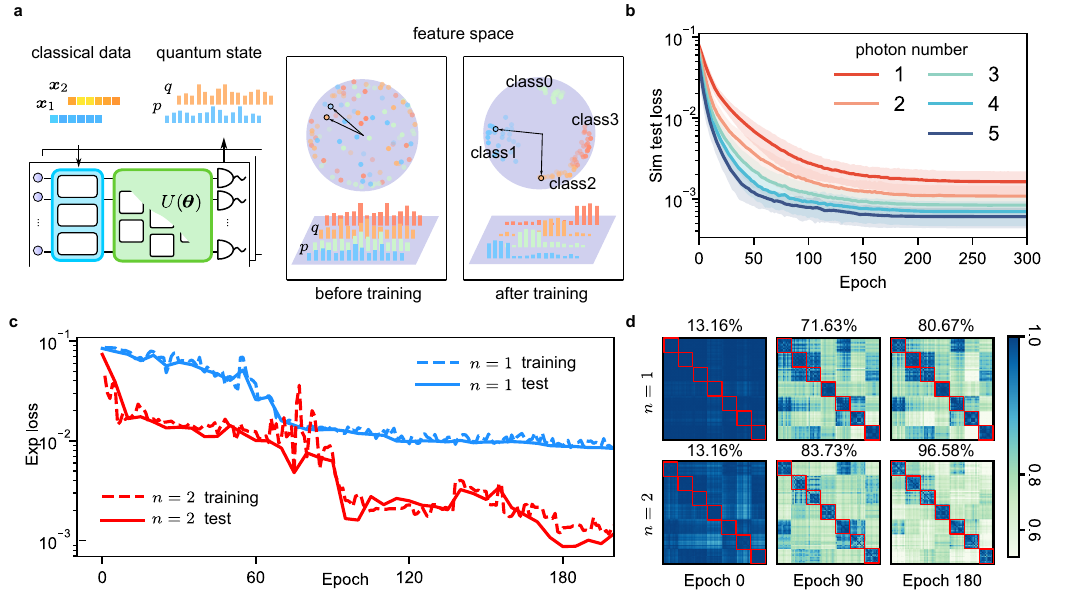}
    \caption{
        \textbf{Experimental quantum metric learning with multi-photon states.}
        \textbf{a}. Two input data points $\vb*{x}_1, \vb*{x}_2$ are encoded into the circuit and processed by a trainable parameterized unitary $U(\vb*{\theta})$. 
        The output probability distributions of photon configurations are measured in the Fock space as $p$ and $q$, respectively. 
        We use a cost function based on the cosine similarity $S_\text{C}(p,q)=\sum_i \sqrt{p_i q_i}$ to guide the learning process, as detailed in Supplementary Note 5. 
        After training, data points from the same class are close to each other on the hypersphere, and points from different classes are well separated.
        \textbf{b}. Simulation of test loss for different photon numbers $n=1,2,3,4,5$.
        For each $n$, the model is trained until convergence with 100 random seeds, and the average loss on the test data is plotted with standard deviation (shaded region).
        \textbf{c}. Experimentally measured training and test loss for single-photon and two-photon cases during the training process. 
        \textbf{d}. Measured Gram matrices on the test data at selected epochs (0, 90 and 180), where darker blue implies higher similarity $S_\text{C}$. 
        The top and bottom rows show the results for single-photon and two-photon experiments, respectively.
        Groups of data from the same classes are highlighted in red boxes.
        Above each matrix we show the pairwise accuracy, which is  defined as the probability of correctly identifying whether a pair of test data belongs to the same or different classes.
    }
    \label{fig:metric}
\end{figure*}

\section{Quantum metric learning with multi-photon states}
\label{sec:metric}

Metric learning~\cite{kaya2019deep} is a widely used supervised learning approach with important applications including face recognition~\cite{schroff2015facenet} and vision language models~\cite{radford2021learning}. 
The goal of metric learning is to learn a distance function or similarity measure in a high-dimensional embedding space, such that similar instances are mapped close together, while dissimilar ones are pushed farther apart. 

The schematic diagram of quantum metric learning with multi-photon states is shown in Fig.~\ref{fig:metric}a.
Input data is encoded into quantum states and transformed by a quantum neural network. 
The resulting output state is measured in the Fock space, yielding a probability distribution $p_\ell(z)=\vert\braket{z}{\psi_\ell(\vb*{\theta})}\vert^2$, where $\ket{z}$ denotes a Fock basis state. 
Given two output probability distributions $\{p_i\}$ and $\{q_i\}$ corresponding to two data points, we use the angular distance between the unit vectors $\sqrt{p_i}$ and $\sqrt{q_i}$ on the hypersphere as the distance measure for metric learning, which is equivalent to computing the cosine similarity $S_\text{C}(p,q)=\sum_i \sqrt{p_i q_i}$. 
The training objective is to minimize the angular distance (i.e., maximize the cosine similarity) between data points from the same class, and vice versa, and is independent of the number of photons or modes (Supplementary Note 5).  

We apply our model to a vowel recognition dataset~\cite{wright2022deep, bandyopadhyay2024single} consisting of vowel-formant frequency vectors from seven classes (see Methods), which we encode into a variational photonic circuit with $m=6$ photon modes. 
First, we conduct numerical simulations for photon numbers $n=1,2,\dots,5$. 
For each configuration, the model is trained with 100 random seeds until convergence, and the average loss values on the test data are plotted in Fig.~\ref{fig:metric}b. 
We observe that test loss values decrease with larger $n$, which is consistent with the growth in learning capacity $R_L(n)$ as a function of $n$, as seen in Eq.~\ref{eq:capacity} and Fig.~\ref{fig:2}c. 

Next, we experimentally implement quantum metric learning using single-photon ($n=1$) and two-photon input states ($n=2$), where in both cases the chip parameters are optimized through online training with the SPSA method from the same initialization.
As shown in Fig.~\ref{fig:metric}c, the training and test losses are substantially lower for two-photon states compared to the single-photon case. 
The Gram matrices, which depict pairwise distances between test data, are shown in Fig.~\ref{fig:metric}d for different training epochs. 
In the plots, test samples are sorted by class. 
For the two-photon case ($n=2$), the matrices exhibit an approximate block-diagonal structure after training: data points from the same class are closer to each other and appear as darker blue blocks (highlighted in red boxes), whereas the off-diagonal blocks correspond to much larger distances between data from different classes and appear in lighter colours.
It is worth noting that for $n=2$, both the average accuracy and class separation improve markedly compared to $n=1$ (Fig.~\ref{fig:metric}d), which is in agreement with both the numerical simulation results (Fig.~\ref{fig:metric}b) and the higher learning capacity of more photons.
More experimental results and implementation details are provided in Supplementary Note 5.

\section{Discussion}
\label{sec:diss}

In this work, we prove that multi-photon states in linear optical circuits provide a fundamentally larger learning capacity, which characterizes the accessible quantum state space and serves as a hallmark for the training and generalization capabilities of QML models~\cite{haug2021capacity,abbas2021power,haug2023generalization,larocca2023TheoryOverparametrization}.
In particular, the unitary learning task pointed out that single-photon states (or coherent states) need $O(m)$ training data, while multi-photon states can accomplish the same tasks with only a constant number of data $O(1)$.
For metric learning, an important machine learning task with broad practical relevance, multi-photon states generalize significantly better by achieving lower loss values, higher test accuracies, and clearer class separation. 
Our experiments, featuring multi-photon online training of two tasks, have shown the feasibility of QML applications on photonic hardware.

Contrasting with existing studies, our work provides a rigorous theoretical understanding for scaling photons in QML, focusing on its impact on training data requirements and generalization performance. 
Our experimental approach is based on readily available and routinely used photonic platforms, involving linear optical circuits, multi-photon states, and single-photon detectors.
Notably, programmable integrated photonic circuits have been commonly applied in classical optical machine learning~\cite{Shen2017, bandyopadhyay2024single}.
Our proposal inspires potential performance gain by upgrading classical instruments such as lasers and photodiodes with multi-photon sources and photon detectors, without altering the internal logic of photonic circuits.

Furthermore, this work not only paves the way for a wide range of photonic QML applications, including variational quantum estimation and quantum simulation~\cite{cimini2024variational, hoch2025quantum, ma2011QuantumSimulation}, but also illuminates quantum computing platforms beyond photons. 
Other bosonic systems such as cold atoms~\cite{young2024AtomicBoson}, microwave photons~\cite{deng2024quantum}, and phonons~\cite{qiao2023SplittingPhonons}, offer controllable numbers of particles that could be engineered for QML tasks.

Finally, we foresee several possible directions for future exploration building on our study.
Firstly, quantum natural gradient methods~\cite{stokes2020quantum} could be investigated to reduce training time by lowering the number of required epochs, as they are inherently compatible with our quantum geometric framework.
Secondly, our protocols can be extended beyond linear optical circuits and photon-number states. For instance, incorporating optical nonlinearity~\cite{Spagnolo,Steinbrecher2019} or combining with continuous-variable states in the non-Gaussian domain~\cite{Andersen2015} could expand the accessible state space and increase the upper bound of the learning capacity $R_L$~\cite{larocca2023TheoryOverparametrization,haug2023generalization}.
Thirdly, while our framework currently targets product states as input, it can be extended to non-separable states, such as those involving entanglement and graph states.
In conclusion, the full potential of photonic QML remains largely untapped, and we believe that leveraging multi-photon states in large-scale programmable photonic systems represents a promising pathway for advancing machine learning.

\section{Acknowledgements}
Y.W. and Z.Y. would like to thank Dr. Aonan Zhang for his valuable discussions.
This research was funded in whole or in part by the Austrian Science Fund (FWF)[10.55776/I5656] (SQOPE), [10.55776/ESP205] (PREQUrSOR), [10.55776/F71] (BeyondC), [10.55776/FG5] (Research Group 5) and [10.55776/I6002] (PhoMemtor). 
For open access purposes, the author has applied a CC BY public copyright license to any author accepted manuscript version arising from this submission.
This project has received funding from the European Union's Horizon 2020 research and innovation programme under grant agreement no. 899368 (EPIQUS), the Marie Skłodowska-Curie grant agreement No 956071 (AppQInfo) and grant agreement no. 101017733 (QuantERA II Programme, project Phomemtor). 
Views and opinions expressed are however those of the author(s) only and do not necessarily reflect those of the European Union or the European Research Council. 
Neither the European Union nor the granting authority can be held responsible for them. 
The financial support by the Austrian Federal Ministry of Labour and Economy, the National Foundation for Research, Technology and Development and the Christian Doppler Research Association is gratefully acknowledged.
The integrated photonic processor was partially fabricated at PoliFAB, the micro- and nanofabrication facility of Politecnico di Milano (\href{https://www.polifab.polimi.it/}{https://www.polifab.polimi.it/}). 
C.P., F.C. and R.O. wish to thank the PoliFAB staff for the valuable technical support. 
R.O. acknowledges financial support by the European Union’s Horizon Europe research and innovation program under QLASS project (Quantum Glass-based Photonic Integrated Circuits, Grant Agreement No. 101135876).

\section{Author contributions}
Y.W. and Z.Y. designed and conducted the experiment.
Y.W., Z.Y. and T.H. developed the theory and algorithms. 
C.P., S.P., A.C. and F.C. conducted the design, fabrication and calibration of the integrated photonic processor. 
R.O. and P.W. supervised the whole project.
All authors discussed the results and reviewed the manuscript.

\bibliographystyle{naturemag}
\bibliography{reference}

\newpage

\section{Methods}

\subsection{Details of the experimental apparatus}
We generate photon pairs at 1550 nm via a spontaneous parametric down-conversion source with a periodically poled potassium titanyl phosphate (ppKTP) crystal which is pumped at 775 nm. 
A tunable delay line is used to adjust the time delay on a single side, compensating for any temporal differences before the photons interfere on the chip.
The integrated photonic circuits utilized in this study are fabricated on a laser-written glass waveguide platform~\cite{pentangelo_high-fidelity_2024, osellame2012FemtosecondLaser, corrielli2021FemtosecondLaser, ceccarelli_low_2020}. 
This circuit features 6 input and output ports and 15 Mach-Zehnder interferometers (MZIs), capable of performing any arbitrary $\text{SU}(6)$ unitary operation. Each MZI is equipped with two thermo-optic resistive phase shifters for independent phase tuning from 0 to $2\pi$. 
Electrical cross-talk between phase shifters is minimized by using a multi-channel current source to control the phase shifters.
In total, 30 phase shifters are required; however, those acting on phase terms placed directly at the output ports (which do not influence quantum interference processes and are not measurable in our apparatus) are not operated in the present experiments.
Superconducting nanowire single-photon detectors are employed to measure single-photon counting events, and a time tagger is used to record coincidence events with a resolution of 15.63 ps.
In the unitary learning task where various learning states are required, we use a programmable optical switch to control the input ports of the photons.

\subsection{Online training on quantum hardware}
We adopt the simultaneous perturbation stochastic approximation (SPSA) method~\cite{spall1992multivariate} for online circuit parameter optimization on our photonic chip. 
SPSA has desirable theoretical properties for noisy intermediate-scale quantum hardware implementations, and has proved successful in experimentally training superconducting quantum platforms~\cite{havlivcek2019supervised, glick2024covariant}. 
By estimating the gradients with only two objective function evaluations per iteration, SPSA avoids linearly increasing computational costs where parameters are varied one at a time. 
Moreover, the gradient approximation process of SPSA is intrinsically noisy, and thus robust to the intractable stochastic fluctuations on noisy quantum hardware.

Specifically, suppose that we want to find the optimal parameter $\bm{x}^*$ to minimize the objective function $f(\bm{x})$. 
For each iteration, we generate a random vector $\Delta$, which is usually drawn from the Rademacher distribution, a discrete probability distribution with a fifty-fifty chance of being $\pm 1$. 
Then the random vector is added and subtracted to the trainable parameters, and $f(\bm{x}+c\Delta)$ and $f(\bm{x}-c\Delta)$ are evaluated through experimental hardware measurements, where $c>0$ is the perturbation amplitude. 
The approximate gradient for $\bm{x}$ is thus given by
\begin{equation*}
    \bm{\delta}_{\text{approx}} = \displaystyle\frac{f(\bm{x}+c\Delta)-f(\bm{x}-c\Delta)}{2c\Delta}.
\end{equation*}
The parameters $\bm{x}$ can be subsequently updated to $\bm{x}-a\bm{\delta}_{\text{approx}}$, where $a>0$ is the learning rate.

The SPSA hyperparameters including $c$ and $a$, can be either estimated by tentative hardware runs or determined by digital simulations. We choose $c=0.4, a=3$ for the task of unitary learning, and $c=0.4, a=150$ for quantum metric learning. These values achieve good performance in digital simulations and show similar convergence behavior in physical experiments. 

\subsection{Learning unitaries with multi-photon states}
To characterize that $U(\vb*{\theta})$ correctly learned the unitary $V$ up to local phases and permutations, we define the matrix closeness $C_\text{M}(U(\vb*{\theta}),V)$ 
\begin{equation}
    C_\text{M}(U,V)=\min_{F,P_\sigma}[1-\frac{1}{m}\text{tr}(V^\dagger F P_\sigma U )]
\end{equation}
where we minimize over the set of local phase unitaries $F(\boldsymbol{\phi})=\sum_{j=1}^m e^{-i\phi_j } \ket{j}\bra{j}$ with angles $\phi_j$ on the $j$th mode and mode permutations $\{P_\sigma\}_\sigma$. Here, $P_\sigma$ are operators that permute modes according to permutation function $\sigma(j)=k$, with $j,k\in\{1,2,\dots,m\}$. For example, for $m=3$, the permutation function with $\sigma(1)=3$, $\sigma(2)=1$ and $\sigma(3)=2$ is a cyclic permutation of modes $(1,2,3)\rightarrow (3,1,2)$.
The matrix closeness has $C_\text{M}(U,V)=0$ only when $U=V$ up to local phases and mode permutations, indicating that we have learned $V$ correctly.

\subsection{Quantum metric learning with multi-photon states}
The vowel dataset in this work has been previously used by multiple studies to demonstrate training novel machine learning hardware setups~\cite{wright2022deep, bandyopadhyay2024single}. 
The dataset consists of 12-dimensional data vectors of formant frequencies extracted from audio recordings, with 7 classes of `ae', `ah', `aw', `er', `ih', `iy' and `uw'.
Each class contains 37 samples, and the total size of the dataset is $37\times 7=259$.
We randomly split the dataset into training and test data with a ratio of $70/30$ and keep the same sample number for all classes. 

In both digital simulations and hardware experiments, we choose a typical variational ansatz where the feature vectors are normalized and encoded onto the phase shifters of the photonic chip (Supplementary Fig. S3). 
For the digital simulations of one to five photons in Fig.~\ref{fig:metric}b, we train using the Adam optimizer~\cite{kingma2014adam} with 100 different random seeds and plot the average simulated test loss curve. 
The learning rate is initially set to 0.1, and is decreased by a factor of 10 when the training loss has stopped decreasing for 10 epochs.
For physical experiments with single-photon and two-photon states, we use SPSA optimization as introduced above. 
All relevant details are provided in Supplementary Note 5. 

\section{Data availability}
All raw experimental data, and the scripts for generating the figures are available at \url{https://github.com/light156/MultiphoQML}.  

\section{Code availability}
All code and necessary data for reproducing this work are available at \url{https://github.com/light156/MultiphoQML}.

\newpage

\end{document}